# A Compact Model of Interface-Type Memristors
## linking physical and device properties

T. F. Tiotto[1,2], A. S. Goossens[1,3], A. E. Dima[2], C. Yakopcic[4], T. Banerjee[1,3], J. P. Borst[1,2], and N. A. Taatgen[1,2]

[1]Groningen Cognitive Systems and Materials Center, University of Groningen, Groningen, The Netherlands
[2]Bernoulli Institute, University of Groningen, Groningen, The Netherlands
[3]Zernike Institute for Advanced Materials, University of Groningen, Groningen, The Netherlands
[4]Department of Electrical and Computer Engineering, University of Dayton, Dayton, OH, USA
{t.f.tiotto,a.s.goossens,t.banerjee,j.p.borst,n.a.taatgen}@rug.nl,
cyakopcic1@udayton.edu

## Abstract

Memristors are an electronic device whose resistance depends on the voltage history that has been applied to its two terminals. Despite its clear advantage as a computational element, a suitable transport model is lacking for the special class of interface-based memristors. Here, we adapt the widely-used Yakopcic compact model by including transport equations relevant to interface-type memristors. This model is able to reproduce the qualitative behaviour measured upon Nb-doped $SrTiO_3$ memristive devices. Our analysis demonstrates a direct correlation between the devices' characteristic parameters and those of our model. The model can clearly identify the charge transport mechanism in different resistive states thus facilitating evaluation of the relevant parameters pertaining to resistive switching in interface-based memristors. One clear application of our study is its ability to inform the design and fabrication of related memristive devices.

## 1 Introduction

One of the greatest challenges faced in the future of computation is in how to extract information from and act upon the ever-increasing quantity of data generated. On one end of the spectrum, the amount of big data being produced is constantly increasing [1] while, on the other, low-power autonomous decision-making is becoming increasingly important in edge computing applications. Creating truly intelligent systems with brain-like functionality has so far been beyond our present-day capabilities; doing so in an energy-efficient manner will be another monumental challenge.

The limitations we face in matching a brain's performance across a variety of tasks stem not only from an algorithmic shortcoming, but also from the comparatively low energy efficiency of our systems. Even our best supercomputers cannot solve many tasks that are natural for us [2]. Nearly all computers are built within the von Neumann framework, which enforces a strict physical and functional division between memory and computation; in most cases this leads to the connection becoming a bottleneck for overall performance, because of the data transfer back and forth during a computation. Given that data creation is accelerating faster than computational capabilities, the von Neumann bottleneck is set to become an ever-greater limiting factor. Relinquishing the von Neumann framework in favour of newer paradigms as neuromorphic computing [3] could pave a way to address both the algorithmic limitations that make our computers unable to be truly intelligent [2] and at the same time enable us to start improving their energy efficiency.

One of the most promising emerging technologies that can be used in starting to emulate the brain is a class of electronic device know as *memristor*. Memristive devices can switch their electrical resistance between two or more levels through the application of external stimuli. They are typically composed of a metal/insulator/metal (MIM) stack which in which microscopic changes occur when an electrical bias is applied across it, resulting in a macroscopic change in resistance. The existence of multilevel conductance states, together with low switching speeds and energy requirements, enables memristors to be used as a remarkably efficient computational substrate which could quite naturally support brain-inspired algorithms and approaches. Memristors can enable the co-location of memory and computation when used to implement synaptic weights in a non-Von Neumann architecture [4].

Different memristive mechanisms are possible depending on the material composition of the MIM cell with phase-change (PCM), filament formation and rupture (ReRAM), magnetic (MRAM), and ferroelectric processes being the most technologically advanced. Interface-type memristors, where electric field controlled resistive switching results from changes occurring at interfaces, are a lesser studied class of memristors but offers distinct advantage for integration in an architecture. Filamentary memristors typically require forming processes before they can be used, which is typically unfavourable for device performance and integration [5, 6]; an attractive feature of interface memristors is that switching behaviour is present in the as-fabricated device. Additionally, they also show reproducible and gradual multi-level resistive switching at room temperature.

It has previously been shown that metal/Nb-doped $SrTiO_3$ (Nb:STO) devices exhibit resistance dynamics in response to repeated voltage pulses describable by a simple power-law of the form $R(n, V) = R_0 + R_1 n^{a+bV}$ [7]. Given that this resistance follows an exponential trajectory, the rate of change slows as its bounds $R_{0/1}$ are approached; such a soft-bounded device can be productively used to model synapses in an ANN while solving life-long learning and catastrophic forgetting issues [8]. While linear synaptic weight changes are often considered ideal, they lead to a response that is either hard-bounded or unbounded; in the continual-learning setting that all biological cognitive systems operate in, this would actually be a disadvantage as it would lead to hindered performance and memory capacity limitations [9]. The resistive switching in Nb:STO Schottky junctions has been documented in literature[10–14] but so far there have been no models that capture the hysteretic current-voltage response.

Large-scale application of memristors is still in its infancy [15] but, like any other component used in computer-aided integrated circuit design, it is paramount to have reliable models of the devices' electrical behaviour, which enable simulation and prediction of the behaviour of both individual memristors as well of the whole integrated circuit. Consequently, a great deal of effort has gone into modelling of the hysteresis curves of these types of memristors: PCM [16–18], ReRAM [19–23], MRAM [24–27], and ferroelectric [28–31]. The Yakopcic model is a generalised, compact representation of the I-V characteristics of a memristor with its core assumption being that a memristor can be represented as two resistors in series, with an internal state variable mixing between the two.

Here we extend the widely-used Yakopcic generalised memristor model [32] to incorporate the often neglected charge transport through metal/insulator interfaces, which is especially important to model systems where the resistive switching occurs at the interface. While the Yakopcic model has been successfully applied to reproducing the electrical behaviour of many memristive devices [33–40], it has not been used to replicate interface-type memristors as the Nb-doped $SrTiO_3$ device we select in this work.

In order to derive a functional description of the resistive switching in interface memristors we modify the electron transmission equations in the Yakopcic model to more closely align with the physical mechanisms relevant for charge transport through Schottky junctions in our Nb:STO device. By fitting this physically-informed compact model to data obtained from real devices fabricated at different sizes we show clear trends in the model parameters, which match the variations expected by our understanding of the underlying physical mechanisms. Given the generality of the model, it could be applied to other interface-type memristors in order to provide insight into their charge transport parameters and help uncover unaccounted physical mechanisms.

## 2 Methods

### 2.1 System Modelled

To model a memristive system, we first have to take account of the relevant underlying transport and switching mechanisms. An important consideration that calls for attention and is often neglected is the role of the metal/insulator interfaces. It can be expected that Schottky barriers form at these interfaces due to the significantly different electronic properties of the contacting materials, which can have a considerable influence on the overall device characteristics. A complete physical device model should include these interfaces to describe the full system. The contact resistance between a metal electrode and oxide is controlled by a space-charge region which forms due to the differences in the Fermi levels of the contacting materials. Typically a Schottky barrier forms at a metal/insulator or metal/semiconductor interface, giving rise to a high contact resistance due to the formation of a space-charge (or depletion) layer in which the mobile carriers are depleted.

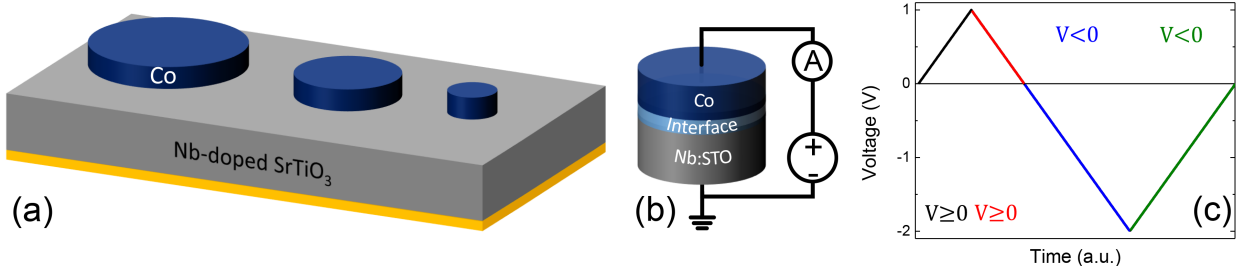

Figure 1: **(a)** Schematic representation of devices used for obtaining experimental data. **(b)** Device structure assumed for modelling, highlighting the interface between the cobalt electrode and the insulating Nb:STO layer. **(c)** Voltage sweeping sequence used to obtain I-V data. The sweep can be split into four parts by considering the polarity (forward bias, $V \geq 0$, or reverse bias, $V < 0$) and the resistance state (LRS $h_1$, or HRS $h_2$).

The effect of such an interface becomes especially important when considering material systems where interface effects play the dominant role in mediating the resistive switching mechanism. This class of memristive devices consist of Schottky interfaces that form between doped wide-band gap oxide, such as SrTiO₃, and high work-function metals. These devices show hysteretic current-voltage characteristics with bipolar switching and large resistance windows together with continuously alterable resistance states. They do not require an initial forming step, as is typically the case for filamentary memristors, which is a significant advantage for practical applications. Both the SET and RESET transitions are gradual and highly modifiable enabling analogue switching between different states within a window. This distinguishes them from the more advanced resistive switching devices such as phase-change materials, in which the RESET operation is abrupt and filamentary memristors which tend to have abrupt SET transitions.

Charge transport is controlled by the Schottky interface and resistive switching is accompanied by changes in the effective Schottky barrier height and width. This is mediated by processes that occur when an electric field is applied: (i) the movement of ionic species, such as oxygen vacancies, and/or (ii) the trapping of electronic charges. In Nb:STO, for example, oxygen vacancy migration under applied electric fields has been observed to be an important factor in mediating the resistive switching [41].

There are various ways in which electrons can be transported across a Schottky interface under bias. Here we consider two of the most important mechanisms which are thermionic emission of electrons over the top of the barrier and quantum-mechanical tunnelling through the barrier. In ideal diodes thermionic emission is the dominant mechanism and this tends to be the most important process in forward bias; in reverse bias, however, tunnelling is expected to strongly contribute to the flow of charge.

There are a vast number of mechanisms through which electrons can traverse the barrier by tunnelling. Some of the widely considered ones include direct tunnelling - which is possible if the barrier is sufficiently thin - or - when the barrier is thicker - through thinner regions of the barrier, for example at higher energies (Fowler-Nordheim) or through defect states (Poole-Frenkel). The relation between the current and voltage depends strongly on system-specific mechanism(s). In this work we consider the theory proposed by Simmons describing the current flow through a generalised barrier [42]. This model approximates tunnelling current through an arbitrary barrier shape by way of a hyperbolic sinusoid function and is hence more widely applicable as it does not assume one specific mechanism.

Here we focus on the interface between Co (work function 5 eV) and Nb:STO (0.1 wt%) (see figure 1(b)). STO is an insulator that becomes an n-type semiconductor when doped with Nb. When Nb:STO is contacted to a metal with a high work function, a Schottky barrier forms at the interface. It has been found that while this doping concentration has no significant effect on the bulk structure and does not induce oxygen vacancies throughout the single-crystalline substrate, it is sufficiently large to make the Nb:STO degenerate. This ensures that the resistance of the semiconductor is relatively low and the Schottky interface contributes the most to the measured resistance.

## 2.2 Measurements

We obtained experimental results from circular cobalt (Co) electrodes of different areas, contacted to an Nb:STO single crystalline substrate with a doping density of 0.1 wt%. The Co contacts were fabricated using a two-step electron beam lithography process using aluminium oxide as an insulation layer to define the contact areas and to prevent electronic cross talk. The bottom of the substrate serves as a back contact for the devices. A device schematic highlighting the different contact areas is shown in figure 1(a).

Two-probe DC voltage sweeps ($0\,\mathrm{V} \rightarrow 1\,\mathrm{V} \rightarrow -2\,\mathrm{V} \rightarrow 0\,\mathrm{V}$) were done from the virgin state; the voltage sweep is shown in figure 1(c). For this study, we measured devices of three different areas and radial dimensions of: $10\,\mu\mathrm{m}$, $32\,\mu\mathrm{m}$, and $100\,\mu\mathrm{m}$. To incorporate device-to-device variation, we obtained results from several devices of each area; the results from individual devices are shown in supplementary figure S1.

## 2.3 Yakopcic Generalised Memristor Model

One model which has seen widespread use is the Yakopcic memristor model [32], which is a generalised, compact mathematical representation of the relationship that exists between input voltage (V) and measured output current (I) in a memristor. The model follows in the steps of [19], in that its fundamental representation of the memristor is as two resistors in series: one with low resistance $R_1$ and the other of high resistance $R_2$. As in [43], this model has been generalised [44] so that the resistors are actually electron transmission equations $h_{1/2}(V)$, making the current a non-linear function of the voltage. Given that a memristor's resistance is analogue and programmable, the two electron transmission equations are weighted and mixed by a dynamic internal state variable $x \in [0,1]$, as proposed by [45]. The internal state variable encapsulates a set of physical variables that influence the material's resistance. Given the known importance of oxygen vacancies, we consider the movement of these ions to be the most relevant internal state. Hence, we model the change in $x$ as non-linear ion motion. Consequently, the I-V curve describing the characteristic relationship between output and input of the device is given by:

$$I(t) = h_1(V(t))x(t) + h_2(V(t))(1 - x(t)) \tag{1}$$

with $h_1$ modelling the behaviour in the low-resistance state (LRS) and $h_2$ in the high-resistance state (HRS).

The dynamics of the internal state variable $x$ are based on two functions $g(V(t))$ (Eq. 2) and $f(x(t))$ (Eq. 3); the former implements the threshold voltage effect by virtue of which $x$ does not change when the voltage drop is sufficiently small, while the latter models the non-linearity of ion motion whereby the change is slowed when approaching its boundaries:

$$g(V(t)) = \begin{cases} A_p\left(e^{V(t)} - e^{V_p}\right), & V(t) > V_p \\ -A_n\left(e^{-V(t)} - e^{V_n}\right), & V(t) < -V_n \\ 0, & -V_n \le V(t) \le V_p \end{cases} \tag{2}$$

$$f(x) = \begin{cases} e^{-\alpha_p(x-x_p)}w_p\left(x, x_p\right), & x \ge x_p \\ 1, & x < x_p \end{cases}$$
$$f(x) = \begin{cases} e^{\alpha_n(x-x_n)}w_n\left(x, x_n\right), & x \le x_n \\ 1, & x > x_n \end{cases} \tag{3}$$

$$w_p\left(x, x_p\right) = \frac{x_p - x}{1 - x_p} + 1$$
$$w_n\left(x, x_n\right) = \frac{x}{x_n} \tag{4}$$

Finally, the state variable dynamics are modelled by integrating the ordinary differential equation (ODE):

$$\frac{dx}{dt} = g(V(t))f(x(t)) \tag{5}$$

### 2.4 Application to Interface Memristors

We expand on the approach presented in [46] by accounting for the fact that in our Nb:STO device different transport mechanisms are dominant in forward and reverse bias [14]. In order to do so, we define the electron transmission equations $h_1$ and $h_2$ (from Eq. 1) as:

$$h_1(V(t)) = \begin{cases} g_{\max,p} \cdot \sinh\left(b_{\max,p} \cdot V(t)\right), & V(t) \geq 0 \\ g_{\max,n} \cdot \left(1 - e^{-b_{\max,n} \cdot V(t)}\right), & V(t) < 0 \end{cases} \tag{6}$$

$$h_2(V(t)) = \begin{cases} g_{\min,p} \cdot \left(1 - e^{-b_{\min,p} \cdot V(t)}\right), & V(t) \geq 0 \\ g_{\min,n} \cdot \sinh\left(b_{\min,n} \cdot V(t)\right), & V(t) < 0 \end{cases} \tag{7}$$

The parameters $g_{\min,n}$, $b_{\min,n}$, $g_{\max,p}$ and $b_{\max,p}$ relate to Schottky transmission and are compact representations of the quantities present in the standard Schottky equation [47]:

$$I(V) = AA^*T^2 e^{-\frac{q\phi_B}{k_B T}} \left(e^{\frac{qV}{nk_B T}} - 1\right) \tag{8}$$

where $A$ is the junction area, $A^*$ the Richardson constant, $q$ the electronic charge, $k_B$ the Boltzmann constant, $T$ the temperature and $\phi_B$ the Schottky barrier height. The ideality factor, $n$ quantifies the deviation from ideal thermionic emission. $g_{\min,p}$, $b_{\min,p}$, $g_{\max,n}$ and $b_{\max,n}$, on the other hand, are related to tunnelling as described by the generalised Simmons model [42]:

$$I(V) = \beta \sinh(\alpha V) \tag{9}$$

where $\alpha$ and $\beta$ are fitting parameters and V is the voltage.

Our choice of $h_1$ and $h_2$ reproduces the model proposed in [45] in that it models the interaction between tunnelling and thermionic emission processes taking place at the interface between metal and insulator (see figure 1(b)). While the ordering of the equations in each bias is not essential, it must be congruent with parameters $x_0$ and $\eta$ in order to correctly capture the dominance of thermionic emission and tunnelling in forward and reverse bias, respectively. Strictly speaking our model does involve the physical parameters of the material, but captures the relative contribution of thermionic and tunnelling in the transport process within the Nb:STO devices.

Figure 1(c) highlights how the model moves between different phases during the voltage sweep used to generate the I-V curves. While being subjected to the voltage sweep shown in figure 1(c), the model moves between using different mixes of thermionic emission and tunnelling as main charge transport mechanism.

An overview of all equations and parameters used by our model is shown table S1.

### 2.5 Parameter Fitting

In order to fit the parameters $\theta$ of our model to experimental data $I$, we simulate its response to the same experimental procedure we used to obtain the I-V characterisation for Nb:STO memristors (section 2.2). We do so by integrating Eq. 5 in order to simulate the memristor models' state evolution in response to the voltage sweeps; at the same time, the current through the device is calculated using Eq. 1. With a slight abuse of notation we can say that each parameter set $\theta$ yields a model which will be in state $x_t$ at each simulation timestep $t$; we can calculate the current $I(x,\theta)$ using Eq. 1 and compare it to the experimentally determined current $I_t$ at time $t$. The objective function used to optimise the fitting is the mean absolute error (MAE):

$$\text{MAE} = \frac{1}{t} \sum_{i=1}^{t} |I(x_t,\theta) - I_t| \tag{10}$$

From the MAE we also derive a mean percent error (MPE) that is useful as a normalised measure of fitting quality between different models:

$$\text{MPE} = 100 * \frac{\text{MAE}}{\sum_{i=1}^{t} |I_t|} \tag{11}$$

This value is normalised with respect to the number of datapoints and consequently any influence from the larger number of reverse bias samples is removed.

We use a two-step automated regression to meaningfully optimise the model so that it can reproduce the currents measured on the Nb:STO devices. The dataset used as basis to derive the models consists of the currents resulting from repeated measurements (meaning a +1/-2 V sweep as detailed in section 2.2 and shown in figure 1(c)) on the memristive devices of different areas (10 µm, 32 µm, and 100 µm area as shown in figure 1(a)). We thus have multiple | Time (s) | Voltage (V) | Current (A) | triples for each of the three devices.

From an initial parameter estimate $\theta_0$ we regress the model parameters $\theta$ upon each available measurement, which are obtained as described in section 2.2. $V_p$, $V_n$ are omitted from the optimisation and set to zero because interface-type memristors do not present a voltage threshold, given that they exhibit analogue modulation of their resistance [14]. Three averaged models are obtained by taking the mean of each parameter value for each device area. The parameters $A_p$, $\alpha_p$, $x_p$ show little to no variation between the three models and are thus excluded from the next regression step by freezing them to the averaged value. The second regression starts from the same initial parameter estimate as the first but ignores $A_p$, $\alpha_p$, $x_p$ - in addition to $V_p$, $V_n$ - for the reasons just mentioned. By grouping by device area the resulting fitted parameters and fitting a Gaussian distribution to each parameter we obtain three models, each one describing the expected I-V relationship for a device fabricated at each area. We can simulate the mismatch in electrical behaviour of different devices by independently sampling parameter sets from the distributions, thus reproducing the device-to-device variation present among the real memristors.

### 2.6 Sensitivity Analysis

Memristors are prone to device mismatch and cycle-to-cycle variation [48]. It is important to understand and characterise the impact that the variations on the underlying physical characteristics have on the electrical behaviour of the devices as circuits and algorithms built upon memristors need to be aware, or at least robust, to these discrepancies in behaviour.

In order to quantify the impact that each parameter has on the overall behaviour of the models we run a sensitivity analysis. In the three models representing a device area, each parameter is increased and decreased until the MPE (Eq. 11) between the current obtained with the model with the varied parameter and the current generated by the original model is 10%. Intuitively: the percentage change in the parameter necessary to see a 10% change in the measured current is in inverse association with the sensitivity of the model to it; the stronger the influence of a parameter, the less it will need to be changed to see a 10% variation in the overall function behaviour of the model.

We exclude $V_p$ and $V_n$ from the analysis as we have elected to fix these to zero when fitting the model, as specified in section 2.5. We also arbitrarily choose to stop the procedure when the change needed in a parameter exceeds 100%, as we consider their impact too small to be relevant after that point.

## 3 Results

### 3.1 Parameter Fitting and I-V simulations

Each parameter in the three models is characterised by a Gaussian distribution whose mean and spread (SD) are reported in table 1. The parameter sets $\theta$ are obtained using the two-step optimisation procedure described in section 2.5.

Figure 2 shows how the simulated I-V curves for the models fitted to the 10 µm, 32 µm, and 100 µm area devices compare to the experimentally measured I-V curves for those memristors. Each model is simulated 100 times with its parameters independently sampled from the Gaussian distribution reported in table 1. The currents collected by measuring the Nb:STO devices are averaged (red curves); the ones resulting from simulating the models display the range of variation upon repeated trials (green curves). The mean absolute error (Eq. 10) and mean percent error (Eq. 11) between the average experimental and simulated I-V curves are presented in table 2.

Table 2: Mean absolute error (Eq. 10) and mean percentage error (Eq. 11) between the average currents measured on the real Nb:STO memristors and those resulting from the simulations of the average models. See figure 2 for a visualisation of the I-V curves.

|     | 10 µm    | 32 µm    | 100 µm   |
| --- | -------- | -------- | -------- |
| MAE | 12.97 A  | 36.76 A  | 64.41 A  |
| MPE | 15.32 %  | 11.46 %  | 9.88 %   |

Table 1: Model parameter values for the three device areas (10 µm, 32 µm, and 100 µm). The mean and standard deviation are derived from averaging and fitting a Gaussian distribution to the model parameters obtained from fitting repeated measurements on devices fabricated with the same area/radius. The bottom section of the table reports parameters that do not vary as a function of device area.

| $\theta$ | 10 µm | | 32 µm | | 100 µm | |
|---|---|---|---|---|---|---|
| | Mean | SD | Mean | SD | Mean | SD |
| $A_n$ | $2.66 \times 10^{-2}$ | $1.70 \times 10^{-3}$ | $2.57 \times 10^{-2}$ | $2.40 \times 10^{-4}$ | $2.43 \times 10^{-2}$ | $1.88 \times 10^{-3}$ |
| $x_n$ | $1.43 \times 10^{-1}$ | 0 | $1.34 \times 10^{-1}$ | 0 | $9.87 \times 10^{-2}$ | 0 |
| $\alpha_n$ | $7.01 \times 10^{-1}$ | $3.75 \times 10^{-1}$ | $2.76 \times 10^{-1}$ | $3.84 \times 10^{-1}$ | $2.50 \times 10^{-1}$ | $2.22 \times 10^{-1}$ |
| $g_{\max,p}$ | $4.34 \times 10^{-4}$ | $1.13 \times 10^{-2}$ | $6.77 \times 10^{-4}$ | $9.69 \times 10^{-3}$ | $6.16 \times 10^{-4}$ | $1.12 \times 10^{-2}$ |
| $b_{\max,p}$ | 4.99 | $1.16 \times 10^{-3}$ | 4.96 | $1.46 \times 10^{-2}$ | 5.19 | $1.15 \times 10^{-2}$ |
| $g_{\max,n}$ | $8.00 \times 10^{-6}$ | $1.27 \times 10^{-6}$ | $4.50 \times 10^{-5}$ | $1.50 \times 10^{-4}$ | $3.30 \times 10^{-5}$ | $4.63 \times 10^{-5}$ |
| $b_{\max,n}$ | 6.27 | $1.35 \times 10^{-1}$ | 5.79 | $1.52 \times 10^{-1}$ | 6.06 | $2.31 \times 10^{-2}$ |
| $g_{\min,p}$ | $3.14 \times 10^{-2}$ | $6.43 \times 10^{-5}$ | $5.99 \times 10^{-2}$ | $3.36 \times 10^{-4}$ | $8.55 \times 10^{-2}$ | $9.26 \times 10^{-5}$ |
| $b_{\min,p}$ | $2.13 \times 10^{-3}$ | $1.40 \times 10^{-1}$ | $3.29 \times 10^{-3}$ | $4.40 \times 10^{-1}$ | $6.88 \times 10^{-2}$ | $9.30 \times 10^{-2}$ |
| $g_{\min,n}$ | $1.50 \times 10^{-5}$ | $9.75 \times 10^{-7}$ | $3.32 \times 10^{-4}$ | $1.25 \times 10^{-5}$ | $1.67 \times 10^{-3}$ | $7.14 \times 10^{-6}$ |
| $b_{\min,n}$ | 3.30 | $3.25 \times 10^{-1}$ | 2.56 | $1.67 \times 10^{-1}$ | 2.12 | $5.54 \times 10^{-2}$ |
| $A_p$ | $7.10 \times 10^{-2}$ | 0 | $7.10 \times 10^{-2}$ | 0 | $7.10 \times 10^{-2}$ | 0 |
| $V_p$ | 0 | 0 | 0 | 0 | 0 | 0 |
| $V_n$ | 0 | 0 | 0 | 0 | 0 | 0 |
| $x_p$ | $1.10 \times 10^{-1}$ | 0 | $1.10 \times 10^{-1}$ | 0 | $1.10 \times 10^{-1}$ | 0 |
| $\alpha_p$ | 9.20 | 0 | 9.20 | 0 | 9.20 | 0 |

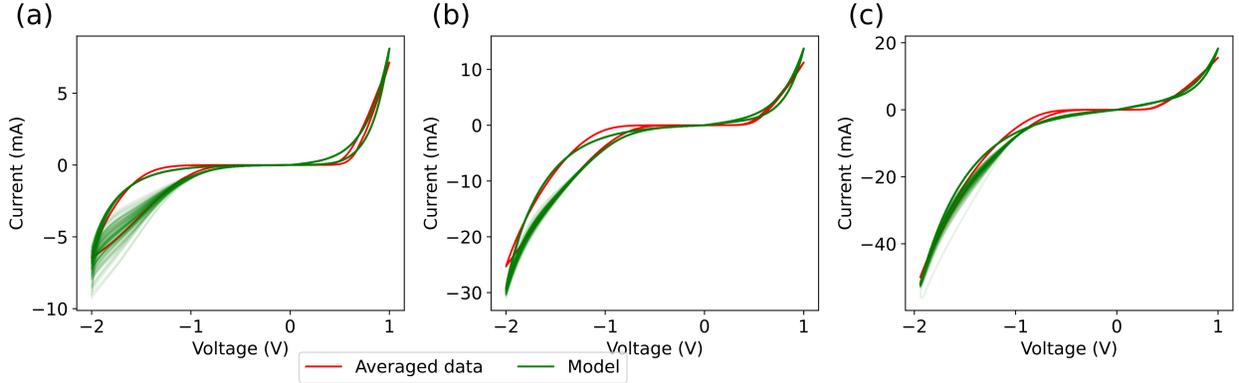

Figure 2: (a) I-V curves of the $10^{-10}$ m² averaged experimental data and of the corresponding models. (b) I-V curve of the $10^{-9}$ m² averaged experimental data and of the corresponding models. (c) I-V curve of the $10^{-8}$ m² averaged experimental data and of the corresponding models. The green curves show the spread of currents resulting from simulating the models 100 times with the parameters independently sampled from the Gaussian distributions reported in table 1. The mean absolute error (Eq. 10) and mean percent error (Eq. 11) between the average experimental currents and those generated by the average models are reported in table 2. The full experimental data can be found in supplementary figure S1.

## 3.2 Relating Model and Device Parameters

In order to start relating model and device parameters we look for trends in the parameters of the average models for the three different areas. While most existing Schottky barrier models do not take device area into consideration, it is known that device edges in finite device areas will affect transport properties. We can gain insight into this rarely considered area variation through our model.

We expect thermionic emission to be mainly present in forward bias while tunnelling is expected to mostly define reverse bias. The resistive switching process occurs under the application of an external electric field due to oxygen vacancy migration and electronic (de)trapping. If the current density is constant, we would expect the current to increase with device area. It has however been observed that in smaller devices the average electric field controlling the charge transport is larger and decreases with increasing device size. This was shown to significantly impact the current-voltage characteristics; notably, smaller devices were shown to have larger resistance windows [49]. We can interpret systematic variations in different model parameters in light of this knowledge and gain more insight into the role of the electric field.

We distinguish two classes of results we can expect from this analysis: (i) trends in model parameters that validate physical mechanisms we are aware of; (ii) correlations in model parameters that let us infer the presence of overlooked physical mechanisms and that will guide future empirical inquiry.

Figure 3 plots the value of parameters for which we have identified correlation with device areas. It can be observed that (a) $A_n$, $\alpha_n$, $b_{\min,n}$ and $x_n$ are negatively correlated to device area, (b) $b_{\min,p}$, $g_{\min,n}$, and $g_{\min,p}$ are positively correlated to area scale; the y-axis is on a log scale so this variation is quite significant. $g_{\min,n}$, $g_{\min,p}$ and $b_{\min,p}$ are independent of area scale and are excluded from the figure.

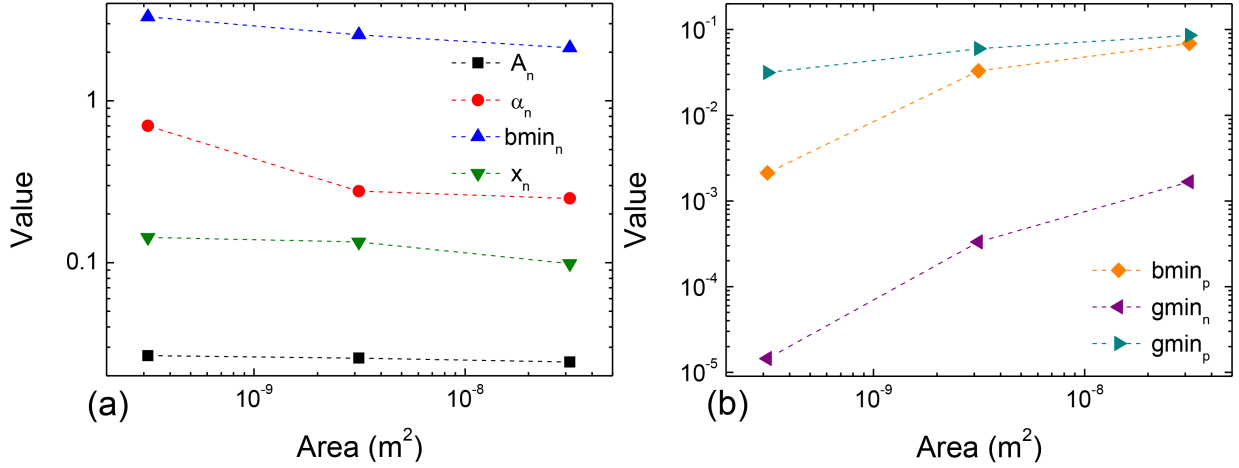

Figure 3: (a) Values of parameters $A_n$, $\alpha_n$, $b_{\min,n}$, and $x_n$ that show a negative correlation with device area. (b) Value of parameters $b_{\min,p}$, $g_{\min,n}$, and $g_{\min,p}$ that are positively correlated to device area.

Given that the parameters used in these equations are strongly tied to the physical mechanisms that govern that transport mechanisms as well as the switching mechanism, we can relate the parameters to device parameters as shown in table 3. This relationship will be further discussed in section 4.

Table 3: Relation between model parameters and device parameters.

| Parameter | Meaning |
| --- | --- |
| $A_p$, $\alpha_p$ | Describe rate of change of the state variable in forward bias. |
| $A_n$, $\alpha_n$ | Describe rate of change of the state variable in reverse bias. |
| $V_n$ | Reverse bias switching threshold voltage |
| $V_p$ | Forward bias switching threshold voltage. |
| $x_p$ | Relates to transition voltage between transport modes in forward bias. |
| $x_n$ | Relates to transition voltage between transport modes in reverse bias. |
| $x_0$, $\eta$ | Capture transition between the different modes of transport. |
| $g_{\max,p}$, $b_{\max,p}$ | Related to tunnelling magnitude (forward bias). |
| $g_{\max,n}$ | Proportional to Schottky barrier height (reverse bias). |
| $b_{\max,n}$ | Proportional to Schottky barrier height (reverse bias). |
| $g_{\min,p}$ | Proportional to Schottky barrier height (forward bias). |
| $b_{\min,p}$ | Inversely proportional to ideality factor (forward bias). |
| $g_{\min,n}$, $b_{\min,n}$ | Related to tunnelling magnitude (reverse bias). |

### 3.3 Sensitivity Analysis

Table 4 reports the results of the sensitivity analysis obtained following the methodology described in section 2.6. Each of the parameters within the average models ("Mean" parameters in table 1) for the three device areas is increased and decreased until a 10 % change in the MPE (Eq. 11) between the original and varied model is observed. The table is organised from most to least impactful parameter in the model, as defined by its average % of variation necessary to see a 10 % change in the MPE across all tested cases. Where a parameter needs to be varied by more than 100 % to see the aforementioned change in MPE we report it as ">100.0", but consider it as 100 % when calculating the averaging as that is the cutoff point for the search.

A preliminary observation of the sensitivity analysis in light of table 3 indicates that the model depends most strongly on the reverse bias parameters and - only in second order - on the forward bias.

Table 4: Sensitivity analysis reporting the amount of change necessary in each parameter to obtain a 10% mean percent error (Eq. 11) change in the current predicted by the model. Results are grouped by the model for each device radius and ordered based on increasing average impact.

| $\theta$ | 10 µm | | 32 µm | | 100 µm | | Average % |
|---|---|---|---|---|---|---|---|
| | Decrease % | Increase % | Decrease % | Increase % | Decrease % | Increase % | |
| $b_{\min,n}$ | 7.7 | 5.8 | 7.7 | 6.1 | 7.8 | 6.5 | 6.93 |
| $g_{\min,n}$ | 22.6 | 22.6 | 18.7 | 18.7 | 16.5 | 16.5 | 19.27 |
| $b_{\max,p}$ | 16.1 | 10.5 | 29.7 | 14.2 | 67.7 | 17.5 | 25.95 |
| $b_{\max,n}$ | 10.7 | 5.7 | 35.0 | 9.6 | >100.0 | 16.6 | 29.60 |
| $A_n$ | 19.4 | 61.6 | 29.2 | >100.0 | 42.8 | >100.0 | 58.83 |
| $g_{\max,p}$ | 36.3 | 36.3 | 58.6 | 58.6 | 90.0 | 90.0 | 61.63 |
| $g_{\min,p}$ | >100.0 | >100.0 | 60.2 | 60.2 | 45.1 | 45.1 | 68.43 |
| $b_{\min,p}$ | >100.0 | >100.0 | 60.3 | 60.7 | 45.4 | 46.0 | 68.73 |
| $A_p$ | 36.7 | 48.8 | 62.1 | >100.0 | 86.8 | >100.0 | 72.40 |
| $g_{\max,n}$ | 47.2 | 47.2 | 81.1 | 81.1 | >100.0 | >100.0 | 76.10 |
| $\alpha_p$ | 40.6 | 99.3 | 56.9 | >100.0 | 74.2 | >100.0 | 78.50 |
| $x_n$ | >100.0 | 33.6 | >100.0 | 64.3 | >100.0 | >100.0 | 82.98 |
| $x_p$ | 60.1 | 55.9 | >100.0 | 92.9 | >100.0 | >100.0 | 84.82 |
| $\alpha_n$ | >100.0 | >100.0 | >100.0 | >100.0 | >100.0 | >100.0 | >100.00 |

## 4 Discussion

In this work we set out to find a physical correlations and interpretation of the parameters in the widely-used Yakopcic memristor model. Starting from empirical data measured on memristive devices fabricated at different scales we defined models able to capture their I-V hysteretic behaviour. We were able to find clear correlations between the models' parameters fitted to the different devices, enabling us to reason back and identify the connection to the underlying physical mechanisms.

### 4.1 Connection Between Model and Physical Device

The parameters $V_p$, $V_n$, $A_p$, $A_n$, $x_p$, $x_n$, $\alpha_p$, and $\alpha_n$ relate to the switching properties of a memristive device. $V_p$ and $V_n$ represent the positive and negative voltages $V(t)$ that need to be exceeded in order to induce a change in the internal state variable and - in turn - the resistance, while factors $A_p$ and $A_n$ control how fast this change is. $x_p$ and $x_n$ are the points where the state variable's motion starts being non-linearly dampened, with $\alpha_p$ and $\alpha_n$ controlling the amount of this damping.

There have been previous attempts to connect the variation in these four pairs of parameters to changes on the filamentary physical device being modelled [44]; it was suggested that $V_p$, $V_n$, $A_p$, $A_n$ most probably vary as a function of the number of oxygen deficiencies, while $x_p$, $x_n$, $\alpha_p$, $\alpha_n$ depend strongly on the degree of crystallinity of the dielectric film and of the electrodes. [44] also looked at the effect of variation of the parameters relating to the I-V equation, namely $a1$, $a2$, and $b$ as defined in the MIM electron transfer

equation:

$$I(t) = \begin{cases} a_1 x(t) \sinh(bV(t)), V(t) \geq 0 \\ a_2 x(t) \sinh(bV(t)), V(t) < 0 \end{cases} \tag{12}$$

$a1$ and $a2$ were found to relate to the thickness of the dielectric layer while $b$ to the proportion of Ohmic-to-tunnelling conduction mechanism.

Looking at figure 3(a) we can appreciate that parameters $A_n$ and $\alpha_n$ are inversely correlated to device area. This results in a more gradual change in the models' state in reverse bias for larger devices, which therefore presents a less abrupt transition from thermionic emission towards tunnelling. Given that tunnelling probability is inversely proportional to the potential barrier width [42] this likely indicates that the smaller the device, the thinner its potential barrier.

Parameter $x_n$ is also negatively correlated to devices' area, with this suggesting that electrons are able to tunnel at lower reverse bias voltages in smaller devices. A higher value for $x_n$ leads to a model where the internal state changes faster. These results are consistent with the expected inverse relation between area and average electric field. The dielectric constant of SrTiO$_3$, $\epsilon_r$, has an inverse relation with the electric field strength[50, 51]. Given that the barrier width $W \propto \sqrt{\epsilon_r}$, a narrower barrier is expected for smaller devices. In reverse bias, the interfacial field is further increased; this is accompanied by a further reduction in the barrier width, facilitating tunnelling at lower voltages.

Turning to look at the reverse bias, we can note that the dual parameters to the ones we have just considered i.e., $A_p$, $\alpha_p$, and $x_p$, are constant with respect to device area. This means that the change in internal state in response to SET pulses is independent of device size. This can be expected for thermionic emission as temperature is an important factor governing at which voltage electrons are able to traverse the barrier, and hence no significant differences are to be expected when comparing the transition to thermionic emission for different devices.

In analysing the parameters that control the magnitudes of the current ($g_{\max,p/n}$, $g_{\min,p/n}$), we have to take care to consider that differences in device area are expected to give rise to larger currents in larger devices.

As can be gleamed from figure 3(b), $g_{\min,p}$ and $b_{\min,p}$ correlate to device area in forward bias. These two parameters define thermionic emission behaviour, which is the dominant transport mechanism in this regime. If we compare the relevant functional form of thermionic emission as we implement it (Eq. 7, $V(t) \geq 0$) to the Schottky emission equation (Eq. 8), we immediately note that $g_{\min,p}$ is proportional to the height of the Schottky barrier ($g_{\min,p} \propto \phi_B$) while $b_{\min,p}$ is inversely proportional to the ideality factor ($b_{\min,p} \propto \frac{1}{\eta}$). This suggests that smaller devices may have a lower barrier height and a higher ideality factor. Both these factors account for the increased current flow observed; a lower barrier would allow more transport through thermionic emission while the larger ideality factor indicates the presence of current flow through other transport mechanisms.

On the converse, $g_{\min,n}$ and $b_{\min,n}$ control electron tunnelling in reverse bias, which is the dominant transport mechanism in this other regime. As can be discerned from figure 3(b), $g_{\min,n}$ is proportional to device area, which would lead us to expect more current to flow through larger devices. This is exactly what we see in figure 2, with a higher current at $-2\,\mathrm{V}$ in the larger devices. However, if we normalise the current flow to account for the larger device area, the current density is larger for smaller devices; this is captured by $b_{\min,n}$ (figure 3(a)), which is inversely proportional to the area.

Any parameter outside those mentioned (i.e., $A_n$, $\alpha_n$, $b_{\min,n}$, $x_n$, $b_{\min,p}$, $g_{\min,n}$, and $g_{\min,p}$) is independent of device area.

## 4.2 Sensitivity Analysis

From Table 4, it can be seen that the model is most sensitive to changes in $b_{\min,n}$ followed by $g_{\min,n}$: the parameters related to reverse bias tunnelling. The third most impactful parameter is $b_{\max,p}$ describes tunnelling in forward bias. From the discussions above and experimental results in [41], tunnelling is largely responsible for the additional current in the LRS compared to the HRS, especially in reverse bias where it is the dominant mechanism. This explains why the model is most sensitive to the parameters controlling the tunnelling behaviour.

Similarly, the model is also highly sensitive to changes in $A_n$, a parameter describing the rate of change of transport from thermionic emission to tunnelling in reverse bias, particularly for the smaller devices. This again alludes to the importance of electronic tunnelling to the switching process and indicates that this mechanism may be more strongly present in smaller devices; this is the probable origin of the larger resistance ratios, which can be appreciated by comparing figure 2(a) to (b) and (c).

The MPE (equation 11) - which we use to evaluate the change in the model's behaviour - is a normalised measure w.r.t. the number of datapoints generated; this means that even though more datapoints are generated in RB due to the deeper voltage sweep, there is not a bias towards the parameters impacting the RB behaviour being reported as more impactful.

### 4.3  Potential Applications and Significance

The model we present in this paper is a first step towards a physical model, as we have started to find correlations between model parameters and corresponding device properties. Defining - and then using - a true physical model would allow us to find an even clearer bijective mapping between the abstract and actual parameter spaces. Moreover, our current model cannot account for many important factors that impact real electrical circuit behaviour as - for example - temperature.

Nonetheless, we can extrapolate some guidelines regarding the expected impact that device design choices may have on the electrical behaviour If the parameters are determined for a particular metal-semiconductor combination, the hysteresis resulting from small variations in experimentally controllable parameters can be simulated. For example, if the simulations indicate the memristive properties can be improved by a change in the Schottky barrier height, experimental strategies can be adopted to achieve this. Specifically, the barrier height could be increased by selecting a metal electrode material with a higher work function, while it could be decreased using a lower work function metal or by introducing a thin oxide barrier in between the metal and semiconductor. Therefore, the model could aid in determining an optimal electrode material or interfacial oxide thickness.

The specific quantitative results of our analyses (as those presented in figure 3 and table 4) are relegated to the properties of Nb:STO and to other interface-type memristors with similar charge transport mechanisms.The model equations presented and our methodology, however, can readily be extended and applied to other types of memristors to include relevant effects that may occur at their metal/insulator interfaces and fully describe the system. Upon defining a physically-informed model of a device it is possible to relate its parameters to the those characterising its physical attributes and/or charge transport mechanisms, for example by running a differential analysis as that we present in section 3.2. After observing how the model parameters change as a function of some device property (in our case, device area) and combining this with a theory of the underlying physical mechanisms, one can define a mapping between the model and device parameter spaces, similarly to what we show in table 3. As mentioned in section 3.2, once this mapping is defined it can productively be used to both verify that predictions made from physical theory are correct and also help in coming up with new theories for the empirical behaviour of the device.

This analysis could also prove valuable in aiding device fabrication itself. Designing and testing electronic components is expensive and time consuming process that may call for a large experimental effort in repeatedly fabricating and testing devices to explore a large design space. Physical theory sets the groundwork for model definition, but its predictions can't be verified without observing the real device's (virtuous loop A in figure 4) or the model's behaviours (virtuous loop B in figure 4). Any of the discrepancies between the model and the device behaviours, and the theory's predictions can be separately used to positively improve the theory itself; integrating the contribution from both model and device can lead to even greater refinements (virtuous loop A+B in figure 4). On the other hand, the model's and fabricated device's behaviours could be integrated to not only inform theory, but each other directly. A sufficiently detailed physical model (or one with a strong mapping to the device parameters) could be used to extrapolate physical parameter sets that promise to yield a device with the desired response; one could then only fabricate devices to those specifications. Model-aided fabrication would therefore be able to inform the creation of the devices themselves, in order to aid in smartly fabricating computational components. It would then be ideal to imagine enacting a virtuous closed-loop procedure (virtuous loop C in figure 4) whereby a physical model would guide fabrication of a device to a certain specification. Model predictions could also be extended to improving physical theory through the results measured on the devices fabricated based on the model's predictions (virtuous loop A+C in figure 4). The measurements on the as-realised device could be compared to the predictions originally made by the model and any error between device response and model predictions could guide optimisation of the latter's parameters. All these relationships could and should be integrated in order to iteratively and synergistically improve theory, model, and physical devices (virtuous loop A+B+C in figure 4). An overview of the relations between physical theory, model, and experimental work are shown in figure 4.

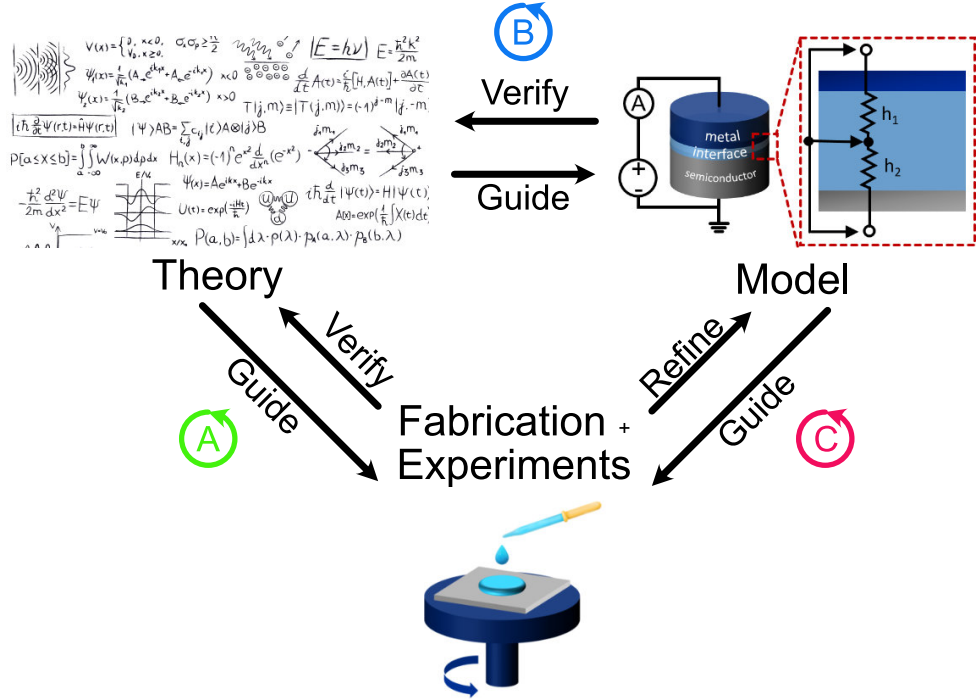

Figure 4: Depiction of the relations defining virtuous loops between physical theory, model, and device fabrication and experiments: A: Physical theory guides fabrication and experiments and is verified by the outcome of these. B: Physical theory guides modelling and is - to a certain degree - verified by its predictions. C: Models can also guide experimentation and device fabrication when a strong enough link exists between the model and device parameter sets. These virtuous closed loops can be combined (A+B, B+C, A+B+C) in a synergistic manner.

## 5  Conclusions

In this paper, we have extended the widely-used Yakopcic memristor model to describe Schottky interfaces, such as those that form at metal-semiconductor/insulator interfaces. Specifically, we consider thermionic emission and tunnelling to be the possible transport mechanisms and the switching is mediated by nonlinear ion motion. We have tested the model on empirical data collected on cobalt/Nb-doped $SrTiO_3$ memristors of different areas and found that the I-V behaviour and hysteresis loop are well-captured in all cases. The model accurately describes the significant differences in transport in forward and reverse bias and is able to capture correlations between the underlying device physics and the model parameters.

These trends are further validation of the physical mechanisms we believe define the electrical characteristics of the devices and also highlight the presence of overlooked physical mechanisms that will guide future empirical inquiry. The generality of the model makes it well-suited for different kinds of interface-type memristors whereby the variables provide insight into device parameters.

## Acknowledgements

T.F.T., A.S.G., A.E.D., J.P.B., T.B. and N.A.T. developed the ideas and wrote the paper. T.F.T., C.Y., and A.E.D. designed the model and ran the simulations. A.S.G. conducted experimental work. T.F.T., A.S.G., C.Y., and A.E.D. derived mathematical results and interpreted the results.

Device fabrication was realised using NanoLab NL facilities. T.F.T. and A.S.G. are supported by the Groningen Cognitive Systems and Materials Centre (CogniGron), University of Groningen.

## Data Availability Statement

The code and modelling data that support the findings of this study are openly available. The experimental data that support the findings of this study are available upon request to the authors.

# Supplementary Data
A Compact Model of Interface-Type Memristors linking physical and device properties

Table S 1: Overview of the Yakopcic model equations containing parameters $\theta$.

| **I-V characteristic** |
|:---:|
| $I(t) = h_1(V(t))x(t) + h_2(V(t))(1 - x(t))$ |
| **LRS equation** |
| $h_1(V(t)) = \begin{cases} g_{\max,p} \cdot \sinh\left(b_{\max,p} \cdot V(t)\right), & V(t) \geq 0 \\ g_{\max,n} \cdot \left(1 - e^{-b_{\max,n} \cdot V(t)}\right), & V(t) < 0 \end{cases}$ |
| **HRS equation** |
| $h_2(V(t)) = \begin{cases} g_{\min,p} \cdot \left(1 - e^{-b_{\min,p} \cdot V(t)}\right), & V(t) \geq 0 \\ g_{\min,n} \cdot \sinh\left(b_{\min,n} \cdot V(t)\right), & V(t) < 0 \end{cases}$ |
| **Threshold voltage** |
| $g(V(t)) = \begin{cases} A_p(e^{V(t)} - e^{V_p}), & V(t) > V_p \\ -A_n(e^{-V(t)} - e^{V_n}), & V(t) < -V_n \\ 0, & -V_n \leq V(t) \leq V_p \end{cases}$ |
| **Non-linear ion motion** |
| $f(x) = \begin{cases} e^{-\alpha_p(x-x_p)}w_p(x,x_p), & x \geq x_p \\ 1, & x < x_p \end{cases}$ |
| $f(x) = \begin{cases} e^{\alpha_n(x-x_n)}w_n(x,x_n), & x \leq x_n \\ 1, & x > x_n \end{cases}$ |
| **Windowing functions** |
| $w_p(x,x_p) = \dfrac{x_p - x}{1 - x_p} + 1$ |
| $w_n(x,x_n) = \dfrac{x}{x_n}$ |
| **State variable dynamics** |
| $\dfrac{dx}{dt} = \eta g(V(t))f(x(t))$ |

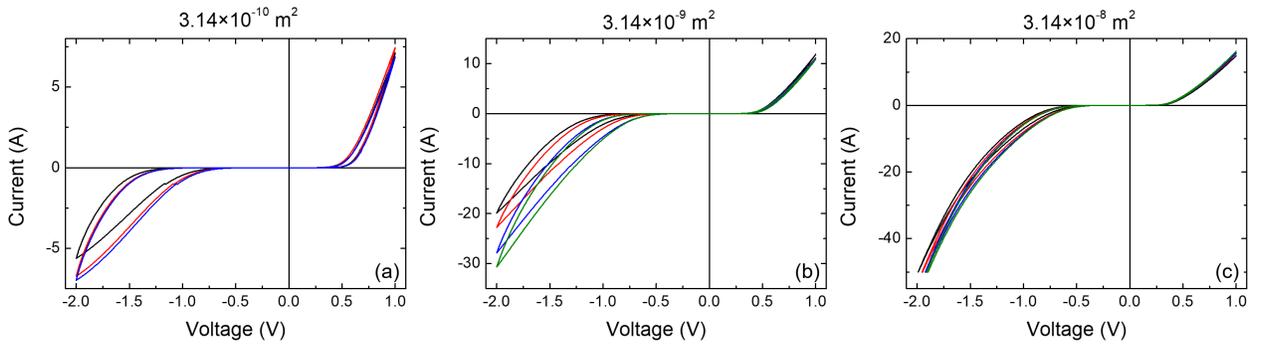

Figure S 1: Current-voltage measurements conducted from voltage sweeps for different devices of areas of (a) 10 µm, (b) 32 µm and (c) 100 µm.

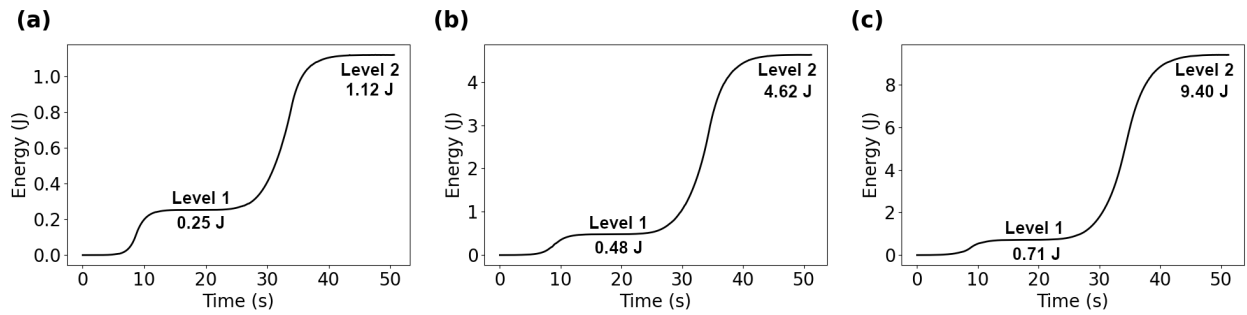

Figure S 2: (a) Energy levels of the 10 μm model over time. (b) Energy levels of the 32 μm model over time. (c) Energy levels of the 100 μm model over time. The two convergence areas, Level 1 and Level 2, correspond to the model switching from HRS to LRS and from LRS to HRS, respectively. Additionally, the figure indicates a positive correlation between device area and the energy for Level 1 as well as Level 2.